\documentclass[useAMS,usenatbib]{mn2e}

\def \nh {$N_{H}$}
\def \chisq {$\chi ^{2}$}

\def\ergs {erg\,s$^{-1}$}
\def\ergscm2 {erg\,s$^{-1}$cm$^{-2}$}

\def\rxj {1RXS\,J170849.0--400910\,}
\def\ea {1E\,2259+586\,}

\def\gtsim{\raisebox{-.5ex}{$\;\stackrel{>}{\sim}\;$}}

\newcommand{\XMM}{{\it XMM--Newton}\,}
\newcommand{\BSAX}{{\it Beppo}SAX\,}
\newcommand{\RXTE}{{\it R}XTE\,}
\newcommand{\INT}{{\it INTEGRAL}\,}

\input psfig.sty

\title[Variability in the AXP \rxj]{Post-glitch variability in the Anomalous X-ray Pulsar
1RXS\,J170849.0$-$400910}

\author[Rea et al.]{N. Rea$^{1,2}$\thanks{Marie Curie Fellow for PhD students to NOVA,  The Netherlands Research School in Astronomy. e-mail: N.Rea@sron.nl}, T. Oosterbroek$^{3}$, S. Zane$^{4}$, R. Turolla$^{5}$,  M. M\'endez$^{1}$, G.L. Israel$^{2}$,
\newauthor L. Stella$^{2}$, F. Haberl$^{7}$  \\
$^{1}$SRON - National Institute for Space Research, Sorbonnelaan 2, 3584 CA, Utrecht, The Netherlands \\
$^{2}$INAF--Astronomical Observatory of Rome, Via Frascati 33,I--00040 Monteporzio Catone (Rome), Italy \\
$^{3}$Science Payload and Advanced Concepts Office, ESA-ESTEC, Postbus 299, NL-2200 AG, Noordwijk, The Netherlands \\
$^{4}$Mullard Space Science Laboratory, University College London, Holmbury St. Mary, Dorking Surrey, RH5 6NT, UK \\
$^{5}$University of Padua, Physics Department, via Marzolo 8, I-35131, Padova, Italy \\
$^{6}$INAF-Istituto di Astrofisica Spaziale e Fisica Cosmica G. Occhialini, Via Bassini 15, I-20133 Milano, Italy \\
$^{7}$Max-Planck-Institut fur extraterrestrische Physik, 85741 Garching, Germany }

\begin{document}

\date{Accepted...Received...}

\pagerange{\pageref{firstpage}--\pageref{lastpage}} \pubyear{2002}

\maketitle

\label{firstpage}

\begin{abstract}

Here we report on the first \XMM\, observation of the Anomalous X-ray
Pulsar \rxj. The source was observed in 2003 August and was found at a
flux level a factor of about two lower than previous
observations. Moreover, a significant spectral evolution appears to be
present, the source exhibiting a much softer spectrum than in the
past. Comparison of the present properties of \rxj\, with those from
archival data shows a clear correlation between the X-ray flux and the
spectral hardness. In particular, the flux and the spectral hardness
reached a maximum level close to the two glitches the source
experienced in 1999 and in 2001, and successively decreased.  Although
the excellent \XMM\, spectral resolution should in principle allow us
to detect the absorption line reported in a phase-resolved spectrum
with \BSAX, and interpreted as a cyclotron feature, we found no
absorption features, neither in the phase averaged spectrum nor in the
phase resolved spectra. We discuss in detail both the possibilities
that the feature in the \BSAX\, data may have resulted from a spurious
detection or that it is real and intrinsically variable. We then
discuss a possible explanation for the glitches and for the softening
of the source emission which followed the flux decrease, in the
framework of the magnetar model.

\end{abstract}


\section{Introduction}

Anomalous X-ray Pulsars (AXPs) and Soft $\gamma$-ray Repeaters (SGRs)
are two peculiar groups of neutron stars (NSs) which stand apart from
other known classes of X-ray sources. They are all radio-quiet,
exhibit X-ray pulsations with spin periods in the $\sim$5--12\,s
range, a large spin-down rate ($\dot{P}\approx
10^{-10}-10^{-13}$s\,s$^{-1})$ and a rather high X-ray luminosity
($L_{X}\approx 10^{34}-10^{36}$\ergs; for a recent review see Woods \&
Thompson 2004). The nature of the X-ray emission from these sources
has been intriguing all along. In fact, for both AXPs and SGRs, the
X-ray luminosity is too high to be produced by the loss of rotational
energy of the neutron star alone, as for more common isolated radio
pulsars.

The magnetic fields of AXPs and SGRs, as estimated from the classical
dipole braking formula $B\sim3.2\times10^{19}\sqrt{P\dot{P}}$~G, are
all above the critical magnetic field at which quantum effects become
important, $B_{QED}\sim 4.4\times10^{13}$~G. At the same time, the
lack of observational signatures of a companion strongly argues
against an accretion powered binary system, favoring instead
scenarios involving isolated NSs. These findings led to the idea that
these two classes of sources should be somehow linked together, and
their X-ray emission related to their very high magnetic fields.


\begin{figure*}
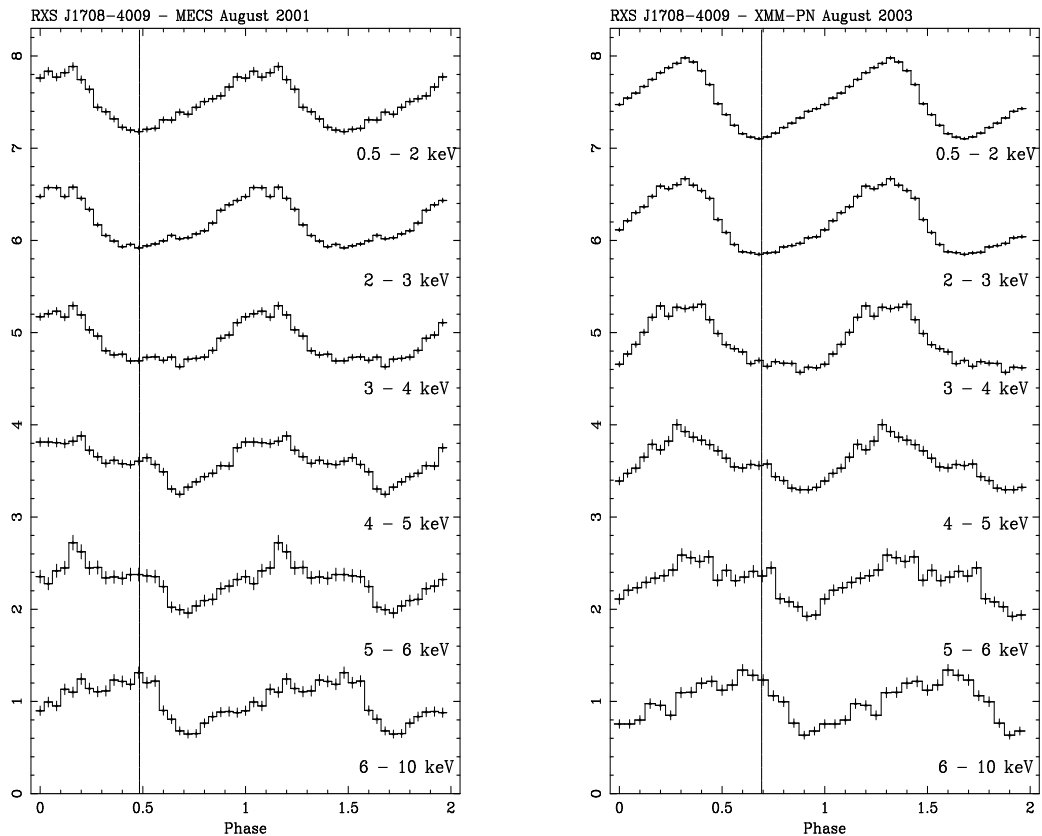

\centerline{
\hbox{ 
\psfig{figure=efold_sax_all_norm_3.ps,width=6cm,height=11cm,angle=-90}
\hspace{1.5cm}
\psfig{figure=efold_all_norm_3.ps,width=6cm,height=11cm,angle=-90}}  }
\caption{Comparison of the folded lightcurves of \rxj\, as a function of energy for the
\BSAX\, (Rea et al. 2003, left) and \XMM\, EPIC-PN (this paper, right) observations. 
Note that due to the time elapsed between the two observations, the
pulse profiles are not phase connected. The solid lines indicate the
minimum of the 0.5-1\,keV in the two cases and is superimposed to the
pictures in order to better show the shift in phase of the two sets of
folded lightcurves at different energies. Units on the y-axis are
counts per seconds with an arbitrary normalization.}
\label{fig:efolding}
\end{figure*}


At present, the model which is most successful in explaining the
peculiar observational properties of AXPs and SGRs is the ``magnetar''
model. In this scenario AXPs and SGRs are thought to be isolated NSs
endowed with ultra-high magnetic fields ($B\sim 10^{14}-10^{15}$\,G)
and their X-ray emission is powered by the magnetic field decay
(Duncan \& Thompson 1992; Thompson \& Duncan 1993, 1995,
1996). Alternative scenarios, invoking accretion from a fossil disk
remnant of the supernova explosion (van Paradijs, Taam \& van den
Heuvel 1995; Chatterjee, Hernquist \& Narayan 2000; Alpar 2001),
encounter increasing difficulties in explaining the data.

While SGRs were first identified through intense, repeated high-energy
bursts observed in the X/$\gamma$-ray band, AXPs were revealed thanks
to their persistent emission in the soft X-rays.  For more than two
decades, AXPs were thought to be steady, soft X-ray emitters and their
X-ray spectrum was well modeled as the superposition of a blackbody
at $kT\sim0.4$\,keV and a steep power-law with photon index
$\Gamma\sim$2--4. No counterparts had been detected at other
wavelengths.  Only in the past few years, mostly thanks to new
generation observatories and to dedicated observational campaigns, our
picture of AXPs has largely changed: 1) a few of them show definite
long term X-ray flux variability on timescales of months (Woods et
al. 2004; Mereghetti et al. 2004; Gavriil \& Kaspi 2004), 2) large
spectral variations with rotational phase were revealed (Israel et
al. 2001; Rea et al. 2003; Tiengo et al. 2005), 3) bursting activity
was detected (Kaspi et al. 2003; Gavriil, Kaspi \& Woods 2002), 4)
glitches were discovered (Kaspi, Lackey \& Chakrabarty 2000, Dall'Osso
et al. 2003; Kaspi \& Gavriil 2003, Morii, Kawai \& Shibazaki 2005),
5) high-energy tails requiring an additional spectral component were
discovered (Kuiper, Hermsen \& M\'endez 2004) and 6) IR/Optical steady
and variable emission was detected (Hulleman, van Kerkwijk \& Kulkarni
2000; Kern \& Martin 2002; Wang \& Chakrabarty 2002; Israel et
al. 2002).


\begin{figure*}
\centerline{\psfig{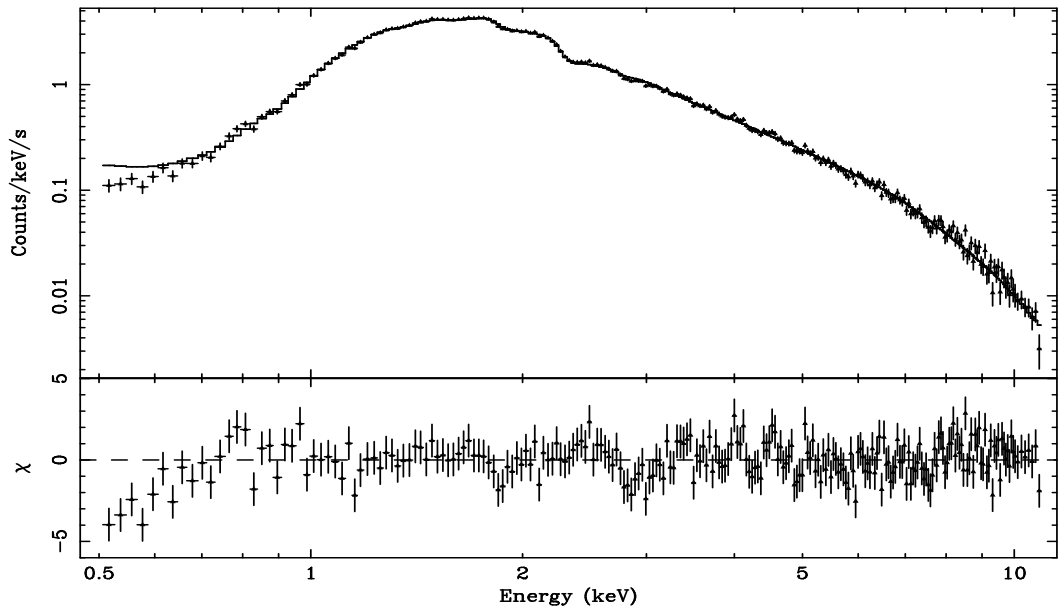}}
\caption{The 0.5--10 keV EPIC-PN phase-average spectrum of \rxj. The residuals 
are referred to a fit with an absorbed blackbody plus a
power-law. Systematic deviations are present in the residuals, most
probably due to instrumental effects.}
\label{fig:phase_averaged_spectrum}
\end{figure*}


\rxj\ was discovered with ROSAT (Voges et al. 1996) and later on a
$\sim$\,11\,s modulation was found in its X-ray flux with ASCA
(Sugizaki et al. 1997).  Early measurements suggested that it was a
fairly stable rotator (Israel et al. 1999). However, in the last four
years the source experienced two glitches, with different post-glitch
recoveries (Kaspi, Lackey \& Chakrabarty 2000, Dall'Osso et al. 2003,
Kaspi \& Gavriil 2003). Searches for optical/IR counterparts ruled out
the presence of a massive companion (Israel et al. 1999), even though
very recently an IR counterpart has been proposed (Israel et al. 2003;
Safi-Harb \& West 2005). A diffuse ($\sim 8^{\prime}$) radio emission
at 1.4 GHz was recently reported, possibly associated with the
supernova remnant G346.5--0.1 (Gaensler et al. 2000).

Pulse phase spectroscopy analysis of two \BSAX\,observations (Israel
et al. 2001; Rea et al. 2003) revealed i) a large spectral variability
with the spin-phase, ii) a strong energy dependence of the pulse
profile shape, and iii) shifts in the pulse phase between the low and
the high energy profiles. High variability of the pulse shape with
energy is now detected at even higher energies, up to $\sim$220~keV,
both with the HEXTE instrument on board of \RXTE\,and the
\INT\, satellite (Kuiper et al. 2005, in preparation).

By analyzing a \BSAX\,observation taken in 2001 (the longest pointing
ever performed on this source), Rea et al. (2003) reported the
presence of an absorption line at $\sim 8$ keV in a phase-resolved
spectrum. The line significance is $\sim4\sigma$.  Interpreting the
feature as a cyclotron line due to resonant scattering yields a
neutron star magnetic field of either $9.2\times10^{11}$\,G or
$1.6\times10^{15}$\,G, in the case of electron or proton
scattering, respectively.

In this paper we report on the first \XMM\,observation of the AXP
\rxj. Timing and spectral data analysis are presented in \S~\ref{obs}. 
The Discussion is divided in two parts: in Section 3 we discuss the
results, readdress the problem of the line significance in the
\BSAX observation, and briefly report on the cyclotron line
variabilities detected so far in other X-ray pulsars; Section 4 is
instead a more theoretical part where we propose an interpretation of
the results in the framework of the magnetar scenario. Conclusions
follow in \S~\ref{conc}.

\begin{figure*}
\centerline{\psfig{figure=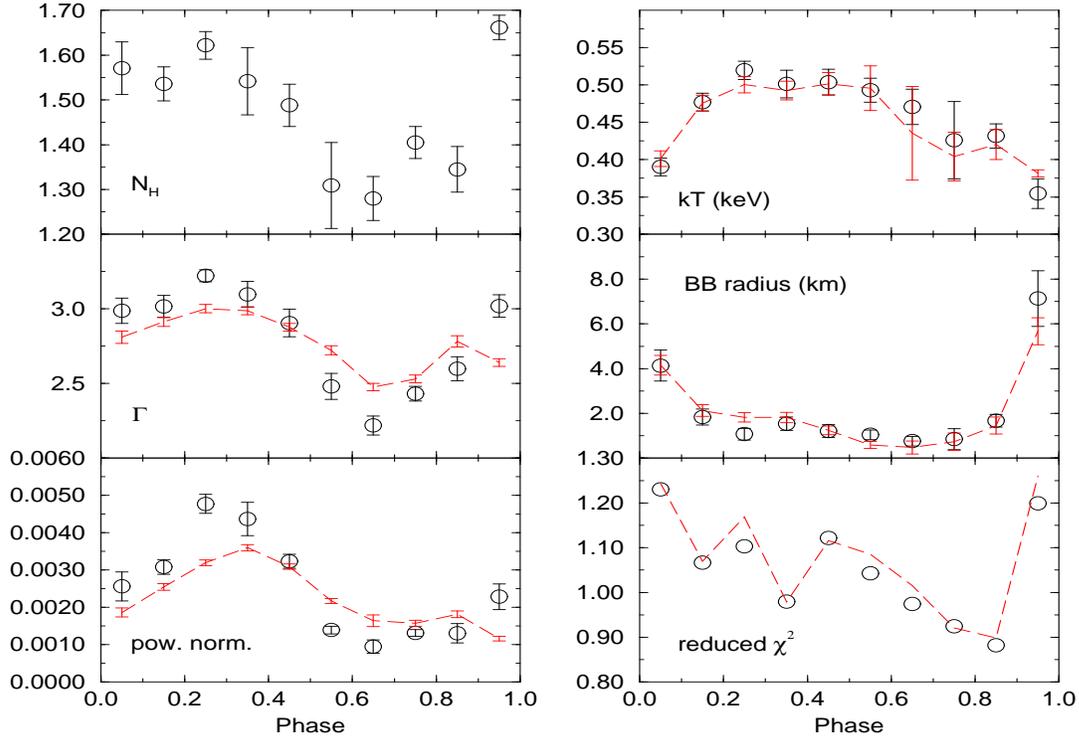,width=15cm,height=10cm,angle=-90}}
\caption{Spectral parameters as a function of phase. The open
circles represent the fits with $N_{H}$ as a free parameter, while the
dashed lines indicate the phase-resolved spectral parameters with $N_{H}$ fixed at
the phase-average value.}
\label{fig:phase_params}
\end{figure*}

\section{Observation and Data Analysis}
\label{obs}

\rxj\, was observed with \XMM\,between 2003 August 28th and 29th,
for $\sim50$~ks. The \XMM\, Observatory (Jansen et al. 2001) includes
three 1500~cm$^2$ X-ray telescopes with an EPIC instrument on each
focus, a Grating Spectrometer (RGS; der Herder et al. 2001) and an
Optical Monitor (Mason et al. 2001). Two of the EPIC imaging
spectrometers use MOS CCDs (Turner et al. 2001) and one uses a PN CCD
(Str\"uder et al. 2001).  The MOS cameras were operated in Prime
Partial Window Mode, while the PN camera was in Prime Small Window
Mode, all with the medium optical photons blocking filter. The Optical
Monitor was operating during the observation but the optical
counterpart of \rxj\,was too faint to be detected. Data were processed
using SAS version 5.4.1. We employed the most updated calibration
files available at the time the reduction was performed (June 2004).
Standard data screening criteria were applied in the extraction of
scientific products. Since a higher background affected the last
$\sim$10\,ks of the observation, we used only the data during
intervals in which the count rate above 10~keV was less than 0.35 counts
s$^{-1}$.  The source events and spectra were extracted within a
circular region of 27$^{\prime\prime}$ centered on the peak of the
point spread function of the source. This non standard radius was used
because the source was located near the edge of the chip. The
background was instead obtained from a source-free region of
27$^{\prime\prime}$. When generating EPIC spectra we have included
events with PATTERN$\leq 4$ (i.e. single and double events). We have
checked that spectra generated with only single events (i.e.
PATTERN$=0$) agreed (apart from normalization factors) with the
spectra generated from single and double events.  All of the EPIC
spectra were rebinned, before fitting, using at least 20 counts per
bin and not oversampling the resolution by more than a factor 3. Since
both MOS and PN give consistent results as far as the timing analysis,
and MOS data are slightly affected by pile-up, in the following we
only report on PN data. We extracted first and second order RGS1 and
RGS2 spectra for both source and background using the standard
procedure reported in the \XMM\,analysis manual but found no evidence
for spectral features.

Thanks to the high timing and spectral PN resolution we have 
been able to perform timing and spectral analysis, as well as Pulse Phase 
Spectroscopy. Results are reported in the following subsections.

\subsection{Timing Analysis}
\label{time} 

In order to determine the spin period of \rxj\,we first barycentered
the events (using the SAS tool {\tt barycen}) and performed a power
spectrum ({\tt powspec}) where we found a periodicity at $\sim$11\,s
(as expected from previous period determinations) composed of a
fundamental and a second harmonic. In order to obtain a precise
estimate of the pulse period, we carried out a period search around
this value ({\tt efsearch}). We obtained a best spin period of
$P_{s}=11.00170\pm0.00004$\,s. The uncertainty was determined by
dividing the observation in 8 intervals and performing a linear fit to
the phase determined in each interval (phase-fitting technique).

The PN phase-folded light curve in different energy bands is shown in 
Fig.\ref{fig:efolding} (right panel), together with that obtained in the 
past by \BSAX. We found that the pulsed 
fraction of the X-ray signal (defined as the 
amplitude of the best-fitting sine wave divided by the, background 
corrected, constant level of the emission) is energy-dependent, 
and it varies from $39.0\pm0.5$\% in the
0.5--2.0\,keV range to $29\pm1.5$\% in the 6.0--10.0\,keV range. These
values are consistent with those reported for the pre-glitches 
\BSAX\,observation (Israel et al. 2001) while both are larger than
those reported for the post-glitches \BSAX\ observation (Rea et al. 2003).

\subsection{Spectral Analysis and Pulse Phase Spectroscopy}
\label{spec}

We generated the PN phase-averaged spectrum and 10 
phase-resolved spectra.  Phase zero was arbitrarily 
defined as the start of the observation (MJD = 52879.9115833) and the 
phase bins are equally spaced. 
The choice of the number of intervals was based on the comparison between
the better quality of the \XMM phase-averaged spectrum with respect to the
\BSAX\, spectrum (for which 6 phase intervals were used) and it was made
prior to the analysis, in order to keep the number of trials to a
minimum in case evidence for a cyclotron line was present.


\begin{figure*}
\centerline{\psfig{figure=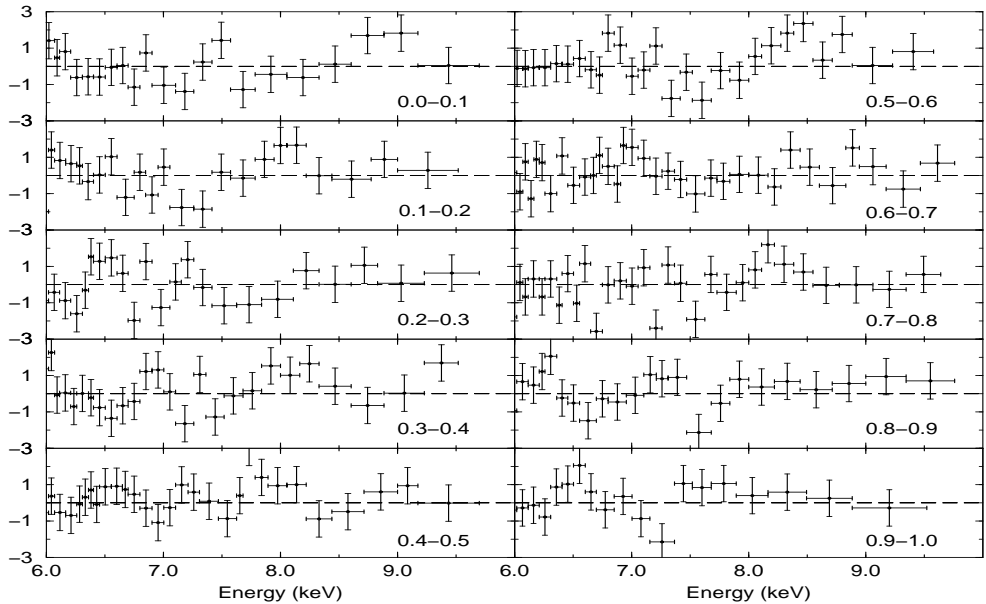,width=13cm,height=8cm,angle=-90}}
\caption{Residuals ($\sigma$) of the fits to the phase-resolved spectrum of \rxj.
Phase intervals are indicated in the lower-right of each panel. This
figure shows the absence of any significant feature during
the whole pulse period.}
\label{fig:phase_residuals}
\end{figure*}


The phase-averaged spectrum in the range 0.5--10\,keV is
satisfactorily fitted by a model consisting of an absorbed blackbody
plus a power law ($\chi_{dof}^2= 1.28$ for 242 degree of freedom, with
a 2\% systematic error included; see
Fig.\ref{fig:phase_averaged_spectrum}).  The best fitting parameters
are: hydrogen column density $ N_{H} = (1.48 \pm0.04)\times
10^{22}$\,cm$^{-2}$, blackbody temperature $kT_{\rm bb}=
0.456\pm0.009$ keV (with a blackbody radius of $R_{\rm bb} =
2.7\pm0.2$~km at a distance of 5\,kpc) and photon index $\Gamma =
2.83\pm^{0.03}_{0.08}$ (all errors are at the 90\,\% confidence
level)\footnote{Note that using the {\em XSPEC} {\it phabs} model
  instead of the {\em wabs} used here, $ N_{H} = (1.36 \pm0.03)\times
  10^{22}$\,cm$^{-2}$ while the other parameters do not change
  significantly.}. The unabsorbed flux in the 0.5--10\,keV range is
$9.1\times 10^{-11}$\,erg\,cm$^{-2}$\,s$^{-1}$ corresponding to a
luminosity of $2.7\times10^{35}$\,erg\,s$^{-1}$ (again for a 5 kpc
distance). In the 0.5--10 keV band, the blackbody component accounts
for $\sim$\,14.9\,\% of the total unabsorbed flux. We found some
deviations from the best fit model in the lower energy range (see
Fig.\ref{fig:phase_averaged_spectrum}), however we believe these
deviations are the result of mixing of spectral shapes, since those
deviations are not present in the phase-resolved spectra and to
calibration issues. Consistently with what has been done in the past
for other AXPs, we tried to fit the spectrum of \rxj\ with two
absorbed blackbodies but this resulted in a worse fit ($\chi_{dof}^2
\sim 2.2$).

The absorbed blackbody plus power law model provides excellent fits
for all the ten pulse phase-resolved spectra, either when leaving \nh\
free or when fixing it to the value obtained from the phase-average
analysis.  The resulting parameters (both for \nh\ fixed and free) are
shown in Fig.~\ref{fig:phase_params}. 
We note that the lowest \chisq\ is observed when the
blackbody flux is lowest, which might suggest that a simple blackbody
is not a very realistic description of the thermal component. As we
tested, by replacing the single temperature blackbody by a disk
blackbody gives a lower $\chi_{dof}^2$ (reduced by up to 0.06) in 9 out of
the 10 phase-resolved spectra.  Of course this does not imply much
about a physical interpretation, but is suggestive of a thermal
emission more complex than a single blackbody.

Note that all the phase-average spectral parameters, but the power-law
photon index $\Gamma$, are consistent (within 1.5$\sigma$ errors) with
the last \BSAX\, observation. We then found here a clear softening in
the spectrum correlated with a decrease of the soft X-ray flux by a
factor of almost two.

As we can see from Fig.\ref{fig:phase_residuals} there is no
significant indication for an absorption line near 8.1 keV, as
reported by Rea et al. (2003). In order to determine an upper limit to
the strength of such a line, we concentrated on the phase interval
which coincides best with that in which the absorption line was found
in the \BSAX\ data. From the re-analysis of the \BSAX\ data, we
noticed that the phase interval in which the absorption line was
strongest is incorrectly stated in the published version of Rea et
al. (2003) (see also \S~\ref{trueornot}).  The correct value is
0.82--1.0 in their phase convention (phase zero corresponding to
minimum of the 0.1--2 keV pulse profile) which corresponds best to
phases 0.5--0.6 and 0.6--0.7 in our phase convention. Thus, it does
not lie in the rising part of the pulse profile, but close to the
minimum in the 0.1--2\,keV range (which actually is the maximum in the
6--10\,keV band; see Fig.\ref{fig:efolding}). We then fitted the
spectrum taken at these phases by adding to the best fitting continuum
(with \nh\ fixed) a {\it cyclabs} model (Mihara et al. 1990; Makishima
et al. 1990) as that used by Rea et al. (2003). We fixed both line
energy and width to 8.1 keV and 0.2 keV, respectively, finding no
significant improvement in the
\chisq\ . We stepped through values of the dimensionless line depth
while the continuum parameters were allowed to change, until a
$\Delta$\chisq\ of 2.706 (90\% uncertainty region for one parameter of
interest) was reached. This occurred at a depth of 0.13 and 0.16 for
the two phase intervals, and we consider the value 0.15 as a 95\%
upper limit to the line depth.

Hints for spectral lines could be found around $\sim7$\,keV in some
phase-resolved spectra and around 1.8\,keV and 2.8\,keV in the
phase-average spectrum (see Fig.\ref{fig:phase_residuals} and
Fig.\ref{fig:phase_averaged_spectrum}) but in all cases the
significance considering one single trial was less than 2$\sigma$,
then highly not statistically significant. We believe them most
probably due to systematic instrumental deviations.

\begin{center}
\begin{figure}

\psfig{figure=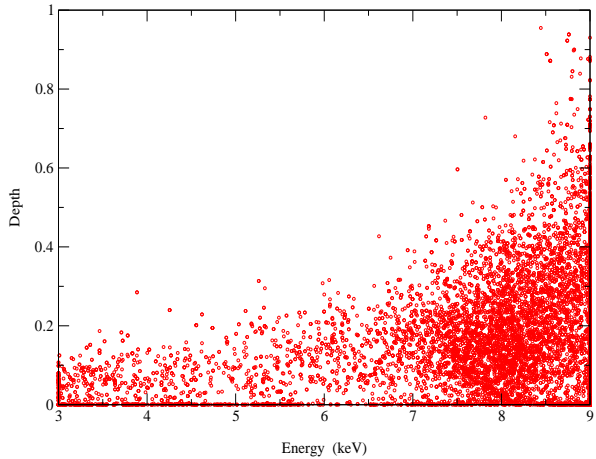,width=10cm,height=7.5cm,angle=-90}
\caption{Monte Carlo simulations. Cyclotron line depth as a function of energy for 10000 simulated spectra.}
\label{fig:simulation}
\end{figure}
\end{center}


\section{Discussion on the results}

In this paper we analyzed the first \XMM\,observation of the AXP
\rxj. We found the source at a flux level 
two times lower than that measured in a previous
\BSAX\ observation (Rea et al. 2003) and with a significantly 
softer spectrum (see Fig.\ref{fig:plot_all} \footnote{Note that the
\BSAX\, flux reported in this plot is slightly lower than one reported in
Rea et al. 2003, this is due to a different, more updated, arf matrix
used here for the re-analysis of this data.} ). The strong phase
dependence of the spectrum, the large pulse shape variability and the
shifts in phase of the pulse profile as a function of energy,
previously detected, are confirmed by the present data. Comparing the
spectral parameters obtained from this \XMM observation with those
derived from the \BSAX \, observation, we find that \nh\ and blackbody
temperature are consistent with being constant, while the blackbody
radius has decreased by about a factor $\sim$2.4. At the same time,
the flux decreased substantially (by a factor 2) and the power-law
index increased from $\sim 2.4$ to $\sim 2.8$. 

The long-term evolution of the source flux and the spectral hardness
are shown in Fig.\ref{fig:plot_all}, where different observations,
spanning nearly five years, have been collected The correlation of the
photon index with the source luminosity is evident by comparing the
two panels. The spectrum became progressively harder as the flux rose
in correspondence of the two glitches and then softened as the
luminosity dropped, following the glitch recovery. This is suggestive
of a scenario in which the mechanism responsible for the glitches is
also at the basis of the enhanced emission and of the spectral
hardening.

The similarity of the second glitch of \rxj\, with the one discovered
during the bursting activity of \ea\ (Kaspi et al. 2003), after which
a similar exponential recovery was seen, suggests that bursts likely
occurred in \rxj\, as well, but the sparse observations did miss them,
as already suggested by Kaspi \& Gavriil (2003). Moreover, the
spectral parameters and the flux changes after the recovery of the
glitch strengthens this idea in comparison with what has been reported
for the post-bursts fading of \ea\,(Woods et al. 2004).


\subsection{The absorption line: Spurious vs. Variability}
\label{trueornot}

As discussed above, we have found no evidence in this \XMM\,
observation for an absorption line at $\sim 8$~keV neither in the
phase-averaged nor in the phase resolved spectra; the upper limit for
the line depth is 0.15 at 95\% confidence level. If we compare this with
the value found by Rea et al. (2003; 0.8$\pm$0.4 at 90\% confidence
level), there is only a very small chance that the two measurements
are consistent. This leaves us with two options to be considered: i)
the \BSAX\, detection is a spurious result, or ii) the absorption line
is variable. 

Before discussing the option of a variable absorption line (see
\S\ref{twist}), we undertook a careful re-analysis of the 
\BSAX\, data. 
This re-analysis resulted in the finding that the phases at which the
absorption line was strongest were given incorrectly in the published
version of Rea et al. (2003), as we noticed earlier: in particular,
the line is strongest close to the pulse minimum in the 0.1-2\,keV
band (or the pulse maximum in the 6-10\,keV band).  We nevertheless
found that the reported estimate of the significance is sound and not
much influenced by different choices in the background subtraction
(annular regions, circular regions far from the source or using blank
field files) or by different extraction regions for the source. The
re-analysis of the \BSAX\, data made varying the extraction radius,
the criterion for the background subtraction and the spectral binning
factor, results in basically the same line properties, which
strengthens our confidence in the robustness of the result. 

Using an F-test method and taking into account the six trials we made
in the phase resolved spectra, we derive a confidence level for the
absorption line of $\sim 4\sigma$.  Note that even if we take into
account all the possible energies at which the feature could lie in
the LECS plus MECS energy range, the confidence level is still $\gtsim
3.5\sigma$. We note that recent works pointed out that the F-test may
be inappropriate in this circumstances (see Protassov et al.~2002),
leading sometimes to incorrect estimates of the significance of a
feature, althought it has been (and still is) widely used to test the
significance of spectral lines. 

In order to further investigate on this issue, we ran a Monte Carlo
simulation of 10000 spectra with only the continuum model (parameters
reported in Rea et al. 2003) and the same number of photons of the
phase resolved spectrum which showed the line in the 2001 \BSAX\,
observation. The results of the simulation is shown in In
Fig.\ref{fig:simulation}. Each circle represents the line depth found in
each of the 10000 spectra, here plotted as a function of the line
energy. From this simulation we found 32 spectra with depth $>$0.8 in
10000 points. We can then reliably say that the probability of the line
being a fluctuation is $<$0.32\%. In summary, we confirm the detection
of the 8.1\,keV absorption line in the \BSAX\, data made by Rea et
al. (2003) at 99.68\% confidence level.

\begin{figure*}
\centerline{\psfig{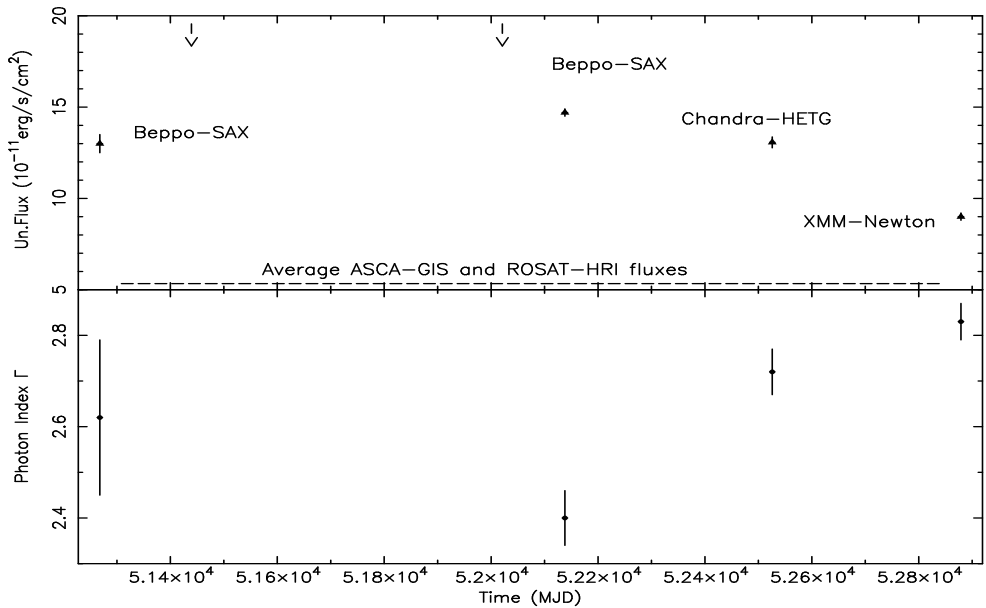}}
\caption{Correlated photon index and X-ray flux changes with time. Dashed line represents 
the average source flux measured by previous ASCA (Israel et al. 1999)
and ROSAT (Sugizaki et al. 1997) observations. The two arrows indicate
the two glitch epochs. All fluxes are unabsorbed and calculated in the
0.5-10\,keV energy band. The Chandra-HETG point (here we report the
first order) comes from the analysis of an unpublished observation
available in the Chandra archive (ObsID 2757).} 

\label{fig:plot_all}
\end{figure*}

The interpretation of the absorption feature as a cyclotron scattering
line proposed by Rea et al.~(2003) was based on the following
criteria: 1) a Gaussian line gives a bad fit and does not reproduce
the asymmetrical shape of the observed feature; 2) the best fitting
model is the {\em XSPEC cyclabs} model; 3) the line strength is highly
phase dependent; 4) no atomic edges or absorption lines are known to
lie around 8.1\,keV (at least without assuming ad hoc shifts possibly
due to the high gravitational redshift or to Zeeman effects in such
strong magnetic field); 5) the relation between the line energy and
width agrees with that of cyclotron scattering features discovered in
other classes of sources (see Fig.\ref{fig:plot_all} in Rea et
al. 2003); 6) the magnetic field inferred from the line energy, either
in the case of an electron or proton cyclotron resonance, is
reasonably consistent with what is expected for a normal neutron star
($\sim10^{12}$\,G) or for a magnetar ($\sim10^{15}$\,G), both being
still open possibilities. Then, if this feature is real, all the above
points hint toward the cyclotron nature of the absorption feature at
8.1\,kev.

Keeping always in mind the possibility that the absorption line in the
\BSAX\, spectrum might be due to statistical fluctuations, in the
following section we discuss a physical mechanisms which could be responsible
for the appearance of a transient cyclotron line in this source in the context of the
magnetar scenario.

\subsection{Other cases of variability of cyclotron absorption features}
\label{linevar}

We conclude this section by summarizing the current status of
detections of variable resonant scattering cyclotron absorption
features in X-ray pulsars, both in binary and in (probably) isolated
systems. 

Cyclotron resonant scattering features were first detected in
accreting X-ray pulsators. Up to now more than ten pulsars in binary
systems are known to exhibit an absorption line in their spectra. In
all these cases, there is unanimous consensus about an interpretation
of the feature in terms of resonant scattering by electrons (see e.g.
Heindl et al. 2004 for a recent review). Variability in the line
energy has been observed in some of these sources, and changes appear
to correlate with the source flux. This behavior finds a quite
natural explanation within the column accretion scenario, in which the
height of the radiation shock depends on the luminosity. At larger
fluxes the shock forms higher up in the column where the (dipole)
field is weaker (Mihara, Makishima \& Nagase 1997; Mihara, Makishima \&
Nagase 2004).

Cyclotron features have been reported so far in the spectra of some
isolated neutron stars too: the soft $\gamma$-repeater SGR 1806-20,
five X-ray dim isolated neutron stars (XDINs) and the radio-quiet
neutron star 1E 1207.4-5209. With the exception of 1E 1207.4-5209
(Bignami et al. 2003), the nature of the line (either electron or
proton) is controversial as yet, and, at least for XDINs, a possible
atomic origin can not be excluded.

The similarity of XDINs pulsational periods ($\sim 3$--10 s), together
with the alleged high values of their magnetic field ($\sim
10^{13}$--$10^{14}$ G), makes comparison with \rxj particularly
interesting. Present data show that absorption features in XDINs are
in all cases dependent on the spin phase. Until recently XDINs were
believed to be steady (albeit some of them are pulsating) soft X-ray
emitters. However, the brighter pulsating member of this class, RX
J0720.4-3125, has been discovered to undergo rather conspicuous
spectral changes over a timescale of a few years (De Vries at
al. 2004; Vink et al. 2004; Haberl et al. 2004). Weather this spectral
evolution is accompanied by significant changes in the absorption line
properties is not completely clear as yet.

Perhaps the most convincing evidence of long-term variations of a
cyclotron line comes from the SGR 1806-20. This source provided the
first possible detection of an absorption feature in a magnetar
candidate (Ibrahim et al. 2002; Ibrahim, Swank \& Parke 2003). The
feature has an equivalent width EW$\sim 0.5$\,keV, width $\sigma\sim
0.2$\,keV and an energy of $\sim 5$ keV. The spectral feature appears
to be transient and has been detected only in the spectrum of a number
of quite energetic bursts. Moreover, recently a hint for a 4\,keV
feature has been found during bursts detected by \XMM\,(Mereghetti et
al. 2005). Despite intensive searches no evidence of any feature has
ever been found in the persistent X-ray emission of this source. The
feature appears to be almost certainly correlated with the onset of
the bursting activity and with periods of increased X-ray luminosity
and spectral hardening.

\section{Interpretation in the magnetar scenario}
\label{twist}

 We discuss here the possibility that the onset of a twist in the
magnetosphere might have been responsible for the two successive glitches, 
for the observed correlation between
spectral hardening, luminosity and glitching activity,  
and, possibly, also for the transient behavior of the 
cyclotron feature.  

Thompson, Lyutikov \& Kulkarni (2002) recently proposed a scenario in
which the magnetars (AXPs and SGRs) differ from standard radio pulsar 
since their magnetic field is globally twisted inside the star, up to a 
strength of about 10 times the external dipole and, at intervals, it can 
twist up the external field. The resulting stresses building up in the
neutron star crust and the magnetic footprints movements can lead to
crustal fractures, known to be responsible of the glitching activity
of the pulsars (Ruderman 1991b; Ruderman, Zhu \& Chen, 1998, Ruderman
2004)\footnote{The strength of the crustal magnetic field can exceed
the critical strength at which anisotropic magnetic field stresses
break the crust. The evolving magnetic field is expected to fracture
the crust at intervals, in a region of order of the coherence field
length (for details see Flowers \& Ruderman 1977; Romani 1990;
Thompson \& Duncan 1993).}.  These fractures could in principle be
violent enough to affect substantially the neutron star interior and
in particular to unpin the crustal neutron superfluid vortex lines
from the crustal lattice, causing the occurrence of a glitch (see
Thompson \& Duncan 1993).

In this picture the vortex lines become unpinned sequentially in small
patches of the crust rather than in the whole crust as in the case
proposed for radio pulsars (Ruderman 1991b). Because of the large
discontinuities of the crustal field, the reconnection of the vortex
unpinned lines could be suppressed at such discontinuities, driving a
pure horizontal field evolution to a point where reconnection is
possible. This, in turn, can cause a second and potentially more
energetic rupture of the crust (Thompson \& Duncan 1993). This
scenario, therefore, can qualitatively explain the occurrence of the
two successive glitches observed from \rxj\, between 1999 and 2001
(Kaspi, Lackey \& Chakrabarty 2001; Dall'Osso et al.  2003; Kaspi \&
Gavriil 2003).  We note that \rxj\, was not totally recovered from the
second glitch at the time of the \BSAX\, observation (Rea et
al. 2003), while it was at time of the \XMM\, observation reported
here.

More interestingly, the onset of a twist in the magnetosphere can in
principle explain both the observed spectral softening and a 
transient appearance of the cyclotron feature. 

In fact, a key feature of twisted, force-free magnetospheres is that
they support current flows.  The presence of charged particle
(electrons and ions) produce both a large resonant scattering depth
and an extra heating of the star surface, the latter induced by
returning currents. Because the electrons distribution is spatially
extended and the resonant frequency depends on the local value of $B$,
repeated scatterings could lead to the formation of a high-energy tail
instead of a narrow line at the cyclotron frequency. In this model,
both the scattering depth (which is strongly dependent on the magnetic
latitude) and the luminosity released by particles hitting the surface
increase with the twist angle.  Therefore, since the spectral hardness
increases with the depth, this implies a positive correlation between
the source X-ray luminosity and the spectral hardening. This is indeed
what the data reported in Fig.\ref{fig:plot_all} indicate.

Moreover, the same model can provide a possible explanation for the
transient appearance of a proton cyclotron feature during the epoch in
which the twist was substantial.  In fact, the charges present in the
magnetosphere are also providing a large optical depth to resonant
proton (ions) cyclotron scattering and the proton resonance occurs
much closer to the star surface than the electron resonance. A
transient absorption feature at the proton cyclotron frequency may
arise then when a) the twist angle is large and b) the X-ray
luminosity at the resonant frequency (evaluated at the star surface,
$\omega_{B,p}$) is large enough to exceed the gravity pull on the
protons, so that positive charges are efficiently confined in a thin
layer close to the star surface (see again Thompson, Lyutikov \&
Kulkarni 2002).


Although not strictly compelling, the correlation between line
appearance, glitching activity and flux enhancement is therefore
intriguing and suggests that the conditions for line formation may
have been met at the epoch of the \BSAX\ pointing.

\section{CONCLUSION}
\label{conc}

Our present analysis of the \XMM\, observation performed on 2003
August of \rxj\, revealed that this source is variable in X-ray over a
period of about five years. This provides further evidence that AXPs,
which were believed to be steady sources for a long time, exhibit both
flux and spectral changes. We pointed out the existence of a clear
correlation between the source flux level and spectral hardening. An
interesting possibility is that both the increased X-ray emission and
the hardening are related to the occurrence of glitches.

The \XMM\, pointing caught the source at a factor of two lower flux
than that of the \BSAX\, observation, performed soon after the last
glitch, and with a significantly softer spectrum. By comparing the
evolution of \rxj\, with that of the AXP \ea, which followed a similar
pattern, it is possible that \rxj\, too had a transient period of
bursting activity, missed in the RXTE monitoring.

We interpret both the occurrence of the glitches and the spectral
softening accompanying the flux decay in the framework of the magnetar
scenario, by the onset of a twist in the external magnetic field
(Thompson, Lyutikov \& Kulkarni 2002). Moreover, the occurrence of the
twist could also account for the possible variability of the cyclotron
line, which was detected only while the source had not entirely
recovered from the last, and more powerful glitch event.

\vspace{1cm}

This paper is based on observations obtained with \XMM, an ESA science
mission with instruments and contributions directly funded by ESA
Member States and the USA (NASA). N.R. thanks L.Kuiper and S.Dall'Osso
for useful discussions about this source, and S. Mereghetti for his
important suggestions and advises on this paper. N.R. is supported by
a Marie Curie Training Grant (HPMT-CT-2001-00245) to NOVA.



\begin{thebibliography}{}

\bibitem {}
   Alpar, M. A., Chau, H.F., Cheng, K.S., Pines, D., 1993, ApJ, 409, 345A  


\bibitem {}
       Alpar, M.A., 2001, ApJ, 554, 12 

 \bibitem {}

Bignami, G.F., Caraveo, P.A., De Luca, A. \& Mereghetti, S., 2003, Nature 423, 725

\bibitem {}
Chatterjee, P., Hernquist, L., \& Narayan, R., 2000, ApJ, 534, 373 



\bibitem {}
     Dall'Osso, S., Israel, G.L., Stella, L., Possenti, A., \& Perozzi, E. 2003, ApJ, 499, 485 

\bibitem {}
     den Herder, J.W. et al., 2001, A\&A, 365, L7  

\bibitem {}
de Vries, C.P., Vink, J., Méndez, M., Verbunt, F., 2004, A\&A 415, L31  


\bibitem {}
       Duncan, R.C., \& Thompson, C. 1992, ApJ, 392, L9 
\bibitem {}


Flowers, E. \& Ruderman, M. A., 1977, ApJ, 215, 302  


\bibitem {}
         Gaensler, B. M., Stappers B. W., Frail, D.A., Moffett, D.A., Johnston, S., Chatterjee, S., 2000, MNRAS, 318, Issue 1, 58 


\bibitem {}
         Gavriil, F.P. \& Kaspi, V.M. 2002, ApJ, 567, 1067G
\bibitem {}
         Gavriil, F.P., Kaspi V.M., \& Woods, P.M. 2002, Nature, 419, 142G 

\bibitem {}
         Gavriil, F.P. \& Kaspi, V.M. 2004, ApJ, 609. L67 


\bibitem {}
         Haberl, F., Zavlin, V.E.,  Trumper, J. \& Burwitz, V., 2004, A\&A 419, 1077 
\bibitem {}
        Haberl 2004, Mem. SAIT 75, 454   

\bibitem {}
Heindl, W.A., et al., 2004, X-ray Timing 2003: Rossi and Beyond. AIP Conference Proceedings, Vol. 714, p.323-330

\bibitem {}
        Ho, W.C.G. \& Lai, D. 2003, ApJ 599, 1293 

\bibitem {}
        Ho, W.C.G. \& Lai, D. 2004, ApJ 607, 420H 

\bibitem {}
   Hulleman, F., van Kerkwijk, M.H. \& Kulkarni, S.R. 2000, Nature, 408, 689


\bibitem{} 

Ibrahim, A.I., Swank, J.H., \& Parke, W. 2003, ApJ, 584, L17  

\bibitem{}

Ibrahim, A.I., Safi-Harb, S., Swank, J.H., Parke, W., Zane, S., Turolla, R., 2002, ApJ, 574, L51   

\bibitem {}
        Israel, G.L., Covino, S., Stella, L., Campana, S., Haberl, F.,
         Mereghetti, S. 1999, ApJ, 518, L107  

\bibitem {}
        Israel, G.L., Oosterbroek, T., Stella, L., Campana, S., Mereghetti, S., Parmar, A., 2001, ApJ, 560, L65 

\bibitem {}
       
         Israel et al., 2002, ApJ, 580, L143 
 \bibitem {}
       
         Israel et al., 2003, ApJ, 589, L93  
\bibitem[2001]{j:01}
        Jansen, F., Lumb, D., Altieri, B., et al. 2001, A\&A, 365, L1  

\bibitem {}
        Kaspi, V.M., Lackey, J.R., Chakrabarty, D. 2000, ApJ, 537, L31 


\bibitem {}         

          Kaspi, V.M. \& Gavriil, F.P. 2003, ApJ, 596, L71 
\bibitem {}

        Kaspi, V.M., Gavriil, F.P., Woods, P.M., Jensen, J.B., Roberts, M.S.E., \& Chakrabarty, D., 2003, ApJ, 588, L93  


\bibitem {}
   
   Kern, B. \& Martin, C. 2002, Nature, 417, 527  
\bibitem {}
        Kuiper,L., Hermsen, W.,  \& M\'endez, M. 2004, ApJ, 613, 1173 


\bibitem {}
        Lai, D., \& Ho, W.C.G., 2003, ApJ, 588, 962

\bibitem {}


        Makishima, K. et al. 1990, ApJ, 365, L59   

\bibitem {}
        Mason, K.O. et al. 2001, A\&A, 365, L36  


\bibitem {}
        Mereghetti, S., Tiengo, A., Stella, L., Israel, G.L., Rea, N., Zane, S., \& Oosterbroek, T., 2004, ApJ, 608, 427 

\bibitem {}
    
         Mereghetti, S., et al. 2005, ApJ in press

\bibitem {}
        Mihara, T., et al., 1990, Nature, 346, 250  


\bibitem {}
 Mihara, T., Makishima, K \& Magase, F., 1997, "All-Sky X-Ray Observations in the Next Decade", 1997, RIKEN, Japan, eds. M. Matsuoka and N. Kawai, p.135  



\bibitem {}
       Mihara, T., Makishima, K \& Magase, F., 2004, ApJ 610, 390  

\bibitem {}
        Morii,M., Kawai, N. \&  Shibazaki. N., 2005, ApJ in press, astro-ph/0412106 


\bibitem { }
        Pavlov, G.G., Sanwal D., Teter, M.A. \& Zavlin, V.E. 2002, AAS Meeting 200, 80.01 
\bibitem[2002]{p:02}
         Protassov, R., van Dyk, D.A., Connors, A., Kashyap V.L., Siemiginowska, A., 2002, ApJ, 571, 545   

\bibitem[2003]{r:03}
        Rea, N., Israel, G.L., Stella, L., Oosterbroek, T., Mereghetti, S., et al. 2003, ApJ, 586, L65 
\bibitem{}

Romani, R.W., 1990, Nature 347, 741
\bibitem{}

Ruderman, M. 1991a, ApJ, 382, 576  

\bibitem{}
Ruderman, M. 1991b, ApJ, 382, 587 

\bibitem{}

Ruderman, M., Zhu, T. \&  Chen, K., 1998, ApJ, 492, 267 
\bibitem{}

Ruderman, M. 2004, Proceedings of the Internation NATO School, Turkey 2004 

\bibitem{}
        Safi-Harb, S. \& West, J., 2005, astro-ph/0501595, Proceedings of the COSPAR Symposium, Paris, July 2004  



\bibitem[2001]{st:01}
         Str\"uder, L. et al. 2001, A\&A, 365, L18  
\bibitem {}
        Sugizaki, M., et al , 1997, PASJ, v.49, p.L25-L30 

\bibitem {}
        Thompson, C., \& Duncan, R.C., 1993, ApJ, 408, 194 
\bibitem {}
        Thompson, C., \& Duncan, R.C. 1995, MNRAS, 275, 255 
\bibitem {}
        Thompson, C., \& Duncan, R.C. 1996, ApJ, 473, 322 

\bibitem {}
 Thompson, C., Lyutikov, M. \& Kulkarni, S.R., 2002, ApJ, 574, 332  

\bibitem {}
        Tiengo, A., Mereghetti, S., Turolla, R., Zane, S., Rea, N., Stella, L. \& Israel, G.L., 2005, A\&A submitted 


\bibitem[2001]{tu:01}
        Turner, M. J. L. et al. 2001, A\&A, 365, L27  

 
\bibitem {}
        van Paradijs, J., Taam, R.E., \& van den Heuvel, E.P.J., 1995, A\&A, 299, L41

\bibitem {}
       Vink, J., de Vries, C.P.; Mendez, M., Verbunt, F., 2004, ApJ, 609, L75 


\bibitem {}
        Voges, W., et al. 1996, IAU Circ., 6420, 2 

\bibitem {}

    Wang, Z. \& Chakrabarty, D. 2002, ApJ, 579, L33 
\bibitem {}
      Woods, P, et al. 2001, ApJ, 552, 748  
\bibitem {}
       Woods, P. \& Thompson, C., 2004, astro-ph/0406133 

\bibitem {}
       Woods, P.M., et al. 2004, ApJ, 605, 378 

\bibitem {}
        Zane, S., Turolla, R., Stella, L., \& Treves, A., 2001, ApJ, 560, 384



\end{thebibliography}
\end{document}